\documentclass[conference]{IEEEtran}

\usepackage{amsmath}
\usepackage{amssymb}
\usepackage[dvips]{graphicx}
\usepackage{epsfig}

\newcommand{\K}{{\sf{K}}}

\newcommand{\ud}{{\mathrm{d}}}

\newcommand{\figsize}{0.4}

\newtheorem{lemma:optphase}{Lemma}
\newtheorem{theo:existence}{Theorem}
\newtheorem{theo:KTcondition}[theo:existence]{Theorem}
\newtheorem{theo:discrete}[theo:existence]{Theorem}

\newtheorem{prop:KTconditionsinglemass}{Proposition}
\newtheorem{cor:R0singlemass}{Corollary}
\newtheorem{prop:Kupperbound}[prop:KTconditionsinglemass]{Proposition}

\begin{document}

\title{Error Probability Analysis of Peaky Signaling over Fading Channels}

%\author{\authorblockN{Michael Shell} \and
%\and \authorblockN{James Kirk\\ and Montgomery Scott}
%\authorblockA{Starfleet Academy\\
%San Francisco, California 96678-2391\\ Telephone: (800)
%555--1212\\ Fax: (888) 555--1212}}

% avoiding spaces at the end of the author lines is not a problem with
% conference papers because we don't use \thanks or \IEEEmembership

\author{\authorblockN{Mustafa Cenk Gursoy}
\authorblockA{Department of Electrical Engineering\\
University of Nebraska-Lincoln, Lincoln, NE 68588\\ Email:
gursoy@engr.unl.edu}}

%\specialpapernotice{(Invited Paper)}

\maketitle

\begin{abstract} \footnote{This work was supported in part by the NSF CAREER Grant
CCF-0546384.} In this paper, the performance of signaling strategies
with high peak-to-average power ratio is analyzed in both coherent
and noncoherent fading channels.  Two recently proposed modulation
schemes, namely on-off binary phase-shift keying and on-off
quaternary phase-shift keying, are considered. For these modulation
formats, the optimal decision rules used at the detector are
identified and analytical expressions for the error probabilities
are obtained. Numerical techniques are employed to compute the error
probabilities. It is concluded that increasing the peakedness of the
signals results in reduced error rates for a given power level and
hence improve the energy efficiency.
\end{abstract}

\section{Introduction}

In wireless communications, when the receiver and transmitter have
only imperfect knowledge of the channel conditions, efficient
transmission strategies have a markedly different structure than
those employed over perfectly known channels. For instance,
Abou-Faycal \emph{et al.} \cite{Abou} studied the noncoherent
Rayleigh fading channel where the receiver and transmitter has no
channel side information, and showed that the capacity-achieving
input amplitude distribution is discrete with a finite number of
mass points. It has also been shown that there always exists a mass
point at the origin. Another key result for noncoherent channels is
the requirement of transmission with high peak-to-average power
ratio in the low signal-to-noise ratio (SNR) regime \cite{Verdu}. In
\cite{gursoy2}, two types of signaling schemes are defined: on-off
binary-shift keying (OOBPSK) and on-off quaternary phase-shift
keying (OOQPSK). These modulations are obtained by overlaying on-off
keying on phase-shift keying. The peakedness of these signals are
controlled by changing the probability of no transmission. OOQPSK is
shown to be an optimally efficient modulation scheme for
transmission over noncoherent Rician fading channels in the low-SNR
regime.

Motivated by the above-mentioned results, we study in this paper the
error performance of using signals with high peak-to-average power
ratios (PAR) over both coherent and noncoherent fading channels.

\section{System Model} \label{sec:model}

We consider the following single-input, single-output flat fading
channel
\begin{gather} \label{eq:model}
y_i = h_i x_i + n_i \quad i = 1, 2, \ldots,
\end{gather}
where $x_i$ is the complex channel input, $y_i$ is the complex
channel output, $h_i$ is the channel fading coefficient, and $n_i$
is the additive Gaussian noise. The fading process $\{h_i\}$ is
assumed to be a stationary, ergodic, and proper complex random
process. Moreover, $\{n_i\}$ is a sequence of independent and
identically distributed (i.i.d.) zero-mean circularly symmetric
complex Gaussian random variables with variance $E\{|n_i|^2\} =
N_0$.

We consider that the transmitter employs OOBPSK or OOQPSK modulation
for transmission. An OOBPSK signal, parametrized by $0 < \nu \le 1$,
has the following constellation points together with their
transmission probabilities:
\begin{align} \label{eq:oobpsk}
x_k = \left\{
\begin{array}{lll}
0 & \text{with prob. } (1-\nu), & k = 0
\\
+\sqrt{\frac{P}{\nu}} & \text{with prob. } (\nu/2), & k = 1
\\
-\sqrt{\frac{P}{\nu}} & \text{with prob. } (\nu/2), & k = 2
\end{array}\right..
\end{align}
Similarly, an OOQPSK signal has the following constellation points
with the corresponding transmission probabilities:
\begin{align} %\label{eq:ooqpsk}
x_k = \left\{
\begin{array}{lll}
0 & \text{with prob. } (1-\nu), & k = 0
\\
\sqrt{\frac{P}{2\nu}}(\pm 1 \pm j) & \text{with prob. } (\nu/4), & k
= 1,2,3,4
\end{array}\right..\nonumber
\end{align}
If the above modulations are adopted, the transmitter either sends
no signal with probability $1-\nu$ or one of the BPSK or QPSK
constellation points with probability $\nu$. Hence, $\nu$ is the
duty cycle of the transmission. Note that these definitions can
straightforwardly be extended to phase-shift keying signals with
larger constellation sizes. Note also that both signals have an
average power of $P$ and a peak power of $\frac{P}{\nu}$, and hence
a peak-to-average power ratio (PAR) of $\frac{1}{\nu}$. Limitations
on the PAR of the signaling scheme may be dictated by regulations or
system component specifications.

\section{Error Probability over Coherent Fading Channels}
\label{sec:coherent}

In this section, we study the error probability of uncoded OOBPSK
and OOQPSK signaling schemes. Here, it is assumed that the receiver
perfectly knows the instantaneous realizations of the fading
coefficients $\{h_k\}$ whereas the transmitter has no such
knowledge.

For the detection of the signals, maximum a posteriori probability
(MAP) decision rule, which minimizes the probability of error, is
employed at the receiver. It is assumed that symbol-by-symbol
detection is performed. Using the property that phase-shift keying
signals have the same energy, the detection rule can be simplified
as follows. The signal $x_k$ for $k \neq 0$ is the detected signal
if
\begin{gather}
\Re(y x_k^*)> \Re(y x_l^*) \quad \forall l \neq k, 0 \quad
\text{and} \quad \Re(y x_k^*) > T \label{eq:smap1}
\end{gather}
where \vspace{-.4cm}
\begin{gather}
T = \frac{1}{2} \frac{|h|P}{\nu} + \frac{N_0}{2 |h|} \ln \left(
\frac{\xi(1-\nu)}{\nu}\right)
\end{gather}
is a threshold value with $\xi = 2$ for OOBPSK signaling and $\xi =
4$ for OOQPSK signaling. In the above formulation, $\Re(z)$ denotes
the real part of the complex scalar $z$, and $z^*$ is the complex
conjugate of $z$. The signal point at the origin, $x_0$, is the
detected signal if $\Re(y x_l^*) < T  \quad \forall l \neq 0$.

\subsection{OOBPSK Signaling} \label{sec:coherentoobpsk}

\begin{figure}
\begin{center}
\includegraphics[width = \figsize\textwidth]{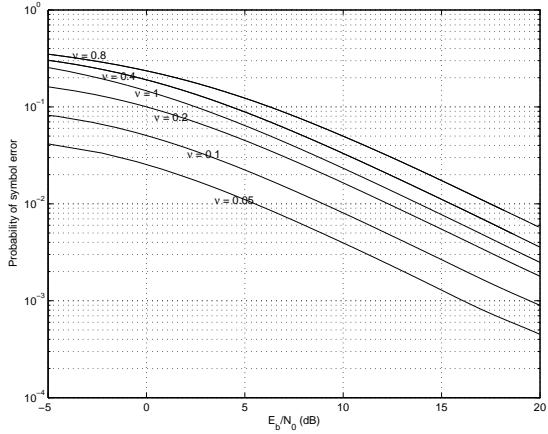}
\caption{Probability of symbol error vs. $E_b/N_0$ for OOBPSK
signaling with different duty factor values, $\nu$, in coherent
Rayleigh fading channels with $E\{|h|^2\} = 1$.}
\label{fig:oobpskrayleigh}
\end{center}
\end{figure}

The error probability of OOBPSK signaling as a function of the
instantaneous realization of the fading coefficient is given by
{\tiny{
\begin{align*}
P_{e | \, |h|} = \left\{
\begin{array}{ll}
 (1\!-\!\nu)2 Q\left( \!T_b\sqrt{\frac{2}{N_0}}
\right)\!+\! \nu Q\left( \sqrt{\frac{2 |h|^2 P}{\nu N_0}} -
T_b\sqrt{\frac{2}{N_0}}\right) & T_b > 0
\\
(1-\nu) + \nu Q\left( \sqrt{\frac{2 |h|^2 P}{\nu N_0}}\right) & T_b
\le 0.
\end{array} \right.
\end{align*}}}
where $Q(\cdot)$ is the Gaussian Q-function and $$T_b = T
\sqrt{\frac{\nu}{P}} = \frac{1}{2}\sqrt{\frac{|h|^2P}{\nu}} +
\frac{1}{2} \sqrt{\frac{N_0^2 \nu}{|h|^2 P}} \ln \left(
\frac{2(1-\nu)}{\nu}\right).$$ The average probability of error is
obtained from
\begin{gather} \label{eq:average}
P_e = \int_0^\infty P_{e| \, |h|} \, d F_{|h|}(|h|)
\end{gather}
where $F_{|h|}$ is the distribution function of the fading
magnitude. Figure \ref{fig:oobpskrayleigh} plots the error
probability curves for OOBPSK signaling in the Rayleigh fading
channel with $E\{|h|^2\} = 1$. It is observed that if the peakedness
of the input signals is increased sufficiently (e.g., $\nu = 0.2,
0.1, 0.05$), significant improvements in error performance are
achieved over ordinary BPSK (i.e., OOBPSK with $\nu =1$)
performance.

\subsection{OOQPSK Signaling}

In this case, it is more convenient to initially obtain the correct
detection probabilities. When $x_1$ is the transmitted signal point,
the correct detection probability is \vspace{-.2cm} {\tiny{
\begin{align}
P_{c|x_1, |h|} = \left\{
\begin{array}{ll}
\left( 1 - Q\left( \sqrt{\frac{|h|^2P}{\nu N_0}}\right)\right)
Q\left( \frac{T_q - \sqrt{\frac{|h|^2P}{2\nu}}}{\sqrt{N_0/2}}\right)
\nonumber
\\
\qquad+\int_0^{T_q} Q\left(\frac{T_q - x -
\sqrt{\frac{|h|^2P}{2\nu}} }{\sqrt{N_0/2}}\right)
\frac{e^{-\frac{\left(x-\sqrt{\frac{|h|^2P}{2\nu}}\right)^2}{N_0}}}{\sqrt{\pi
N_0}} \, \ud x & T_q > 0
\\
\left( 1 - Q\left( \sqrt{\frac{|h|^2P}{\nu N_0}}\right)\right)^2 &
T_q \le 0
\end{array}\right.
\end{align}}}
where
\begin{gather}
T_q = T\sqrt{\frac{2 \nu}{P}} = \sqrt{\frac{|h|^2 P}{2\nu}} +
\sqrt{\frac{N_0^2 \nu}{2 |h|^2 P}} \ln\left(
\frac{4(1-\nu)}{\nu}\right).
\end{gather}

\begin{figure}
\begin{center}
\includegraphics[width = \figsize\textwidth]{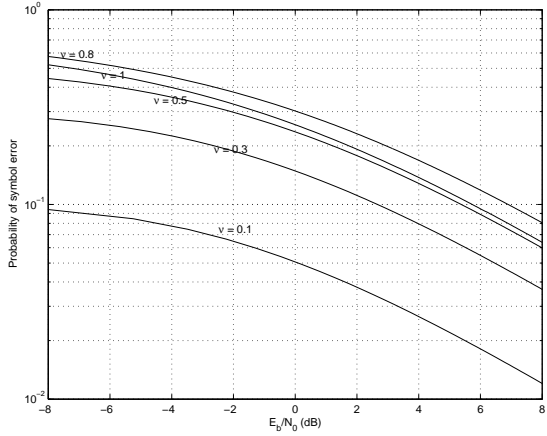}
\caption{Probability of symbol error vs. $E_b/N_0$ for OOQPSK
signaling with different duty factor values, $\nu$, in coherent
Rayleigh fading channels with $E\{|h|^2\} = 1$.}
\label{fig:ooqpskrayleigh}
\end{center}
\end{figure}

If $x_0$ is sent, the correct detection probability is
\begin{gather}
P_{c | x_0, |h| } = \left\{
\begin{array}{ll}
 4\int_0^{T_q} \left( \frac{1}{2} -
Q\left( \frac{T_q - x}{\sqrt{N_0/2}} \right)\right)
\frac{e^{-\frac{x^2}{N_0}}}{\sqrt{\pi N_0}} \, \ud x & \text{if }T_q
> 0
\\
0 & \text{if } T_q \le 0
\end{array} \right.. \nonumber
\end{gather}
Now, the error probability as a function of the fading coefficients
and the average error probability are given by
\begin{align}
P_{e| \, |h|} &= 1 - ((1-\nu) P_{c | x_0, |h|} + \nu P_{c | x_1 ,
|h|} ),
\end{align}
and \vspace{-.4cm}
\begin{align}
 P_e &= \int_0^\infty P_{e| \, |h|} \, d F_{|h|}(|h|),
\end{align}
respectively, where $F_{|h|}$ is the distribution function of the
fading magnitude. Fig. \ref{fig:ooqpskrayleigh} plots the error
probability curves for OOQPSK signaling again as a function of the
SNR per bit (i.e., SNR normalized by the entropy of OOQPSK signals)
in the coherent Rayleigh fading channel with $E\{|h|^2\} = 1$. It is
seen that error performance improves if $\nu \le 0.5$ and
significant gains are obtained when $\nu = 0.1$ which reguires a
10-fold increase in the peak-to-average power ratio when compared to
ordinary QPSK signaling. It is interesting to note that if $\nu <
0.8$, the threshold value $T_q \to \infty$ as $P\to 0$. Therefore,
for sufficiently small $P$, the zero signal, $x_0$, is always chosen
as the detected signal and the error probability becomes $\nu$.
Indeed, it is observed in Fig. \ref{fig:ooqpskrayleigh} that the
error curves corresponding to OOQPSK signaling with $\nu < 0.8$
approach to $\nu$ as SNR decreases. If $\nu > 0.8$, $T_q \to
-\infty$ as $P \to 0$. When $T_q \le 0$, the zero signal is never
detected. Note that this behavior is also exhibited if OOBPSK
modulation is used.

\section{Error Probability over Noncoherent Fading Channels}
\label{sec:noncoherent}

In this section, we consider the scenario in which neither the
transmitter nor the receiver knows the instantaneous realizations of
the fading coefficients. We consider a fast Rician fading
environment and hence $\{h_k\}$ is a sequence of i.i.d. proper
complex Gaussian random variables with mean $E\{h_k\} = m$ and
variance $E\{|h_k|^2\} = \gamma^2$. It is assumed that channel
statistics and hence the values of $m$ and $\gamma^2$ are assumed to
be known at the transmitter and receiver. The conditional output
probability density function given the input is
\begin{gather} \label{eq:cpdfn}
f_{y|x}(y|x) = \frac{1}{\pi (\gamma^2 |x|^2 + N_0)} e^{-\frac{|y -
mx|^2}{\gamma^2 |x|^2 + N_0}}.
\end{gather}
The MAP detection yields
the following decision rule: $x_k$ for $k \neq 0$ is the detected
signal if
\begin{align}
\Re(y x_k^*) > \Re(y x_l^*) \quad \forall l \neq k, 0
\label{eq:smapn1}
\end{align}
and
\begin{align}
 \frac{\gamma^2 P}{\nu} \, |y|^2 + 2 N_0 |m| \,
\Re(y x_k^*)
> T \label{eq:smapn2}
\end{align}
where {\small{
\begin{gather} \label{eq:thresholdn}
T = N_0 \frac{P}{\nu} |m|^2 + N_0 \left( \gamma^2 \frac{P}{\nu} +
N_0\right) \ln\left( \frac{\xi (1-\nu)}{\nu} \left( \gamma^2
\frac{P}{\nu N_0} + 1\right)\right)
\end{gather}}}
with $\xi = 2$ for OOBPSK signaling and $\xi =4$ for OOQPSK
signaling. The signal point $x_0$ is the detected signal if
\begin{gather} \label{smapn0}
\frac{\gamma^2 P}{\nu} \, |y|^2 + 2 N_0 |m| \, \Re(y x_l^*) < T
\quad \forall l \neq 0.
\end{gather}
For brevity, we only discuss OOQPSK signaling in the following. The
analysis of the OOBPSK signaling can be found in the extended
version of this paper \cite{gursoy3}.

\subsection{OOQPSK Signaling}

\begin{figure}
\begin{center}
\includegraphics[width = \figsize\textwidth]{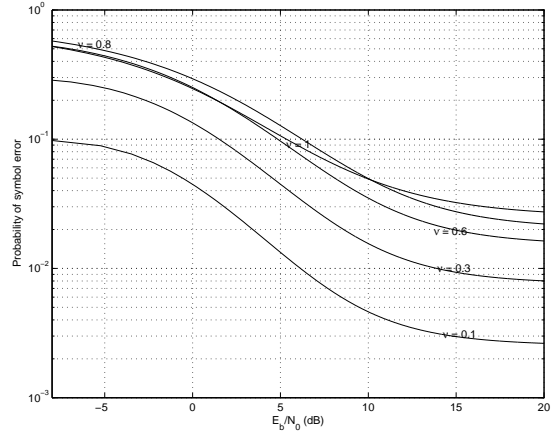}
\caption{Probability of symbol error vs. $E_b/N_0$ for OOQPSK
signaling with various duty factor values, $\nu$, in the noncoherent
Rician fading channel with Rician factor $\K = |m|^2 / \gamma^2 =
5$.} \label{fig:ooqpskriciannk5}
\end{center}
\end{figure}

As before, we first express the correct detection probabilities. If
a nonzero signal point is sent and $T>0$, we have {\tiny{
\begin{align*}
P_{c|x_1} &= \left( 1 - Q\left( \frac{
\sqrt{\frac{|m|^2P}{2\nu}}}{\sqrt{\frac{\gamma^2P}{2\nu}+\frac{N_0}{2}}}\right)\right)Q\left(
\frac{D_q -
\sqrt{\frac{|m|^2P}{2\nu}}}{\sqrt{\frac{\gamma^2P}{2\nu}+\frac{N_0}{2}}}\right)
\nonumber
\\
&\quad +\int_0^{D_q} Q\left(\frac{\sqrt{C_q-(x+A_q)^2}-A_q
-\sqrt{\frac{|m|^2P}{2\nu}}
}{\sqrt{\frac{\gamma^2P}{2\nu}+\frac{N_0}{2}}}\right)
\nonumber \\
& \qquad \times
\frac{e^{-\frac{\left(x-\sqrt{\frac{|m|^2P}{2\nu}}\right)^2}{\frac{\gamma^2P}{\nu}+N_0}}}{\sqrt{\pi
(\frac{\gamma^2P}{\nu}+N_0)}} \, \ud x \label{eq:Peqn1}
\end{align*}}}
where {\small{
\begin{align}
A_q = \frac{N_0 |m|}{\gamma^2} \sqrt{\frac{\nu}{2P}}, \quad  C_q = T
\frac{\nu}{\gamma^2P} + 2A_q^2, \quad D_q = \sqrt{C_q - A_q^2} - A_q
\end{align}}}
with $T$ given in (\ref{eq:thresholdn}). If $T < 0$,
\begin{gather} \label{eq:Peqn1T0}
P_{c|x_1} = \left(1 - Q\left(
\frac{\sqrt{\frac{|m|^2P}{2\nu}}}{\sqrt{\frac{\gamma^2P}{2\nu}+\frac{N_0}{2}}}\right)\right)^2.
\end{gather}
If $x_0$ is the transmitted signal, {\footnotesize{
\begin{align}
P_{c | x_0} = \left\{
\begin{array}{ll}
 \!\!\!\!4\int_0^{D_q} \left( \frac{1}{2} -
Q\left( \frac{\sqrt{C_q - (x+A_q)^2}-A_q}{\sqrt{\frac{N_0}{2}}}
\right)\right) \frac{e^{-\frac{x^2}{N_0}}}{\sqrt{\pi N_0}} \, \ud x
& \!\!\text{if }T
> 0
\\
\!\!\!\!0 & \!\!\text{if } T \le 0
\end{array} \right.. \nonumber
\end{align}}}
Finally, the error probability is obtained from
\begin{gather}
P_e = 1 - ((1-\nu)P_{c|x_0} + \nu P_{c|x_1}).
\end{gather}
Fig. \ref{fig:ooqpskriciannk5} plots the error probability curves
for OOQPSK signaling as a function of SNR per bit in the noncoherent
Rician fading channel with Rician factor $\K = |m|^2/\gamma^2 = 5$.
Similarly as before, it is seen that error performance improves for
sufficiently small duty factors. It is again noted that for $\nu <
0.8$, $C_q \to \infty$ as $P \to 0$. Hence, for sufficiently small
$P$, the zero signal is always declared as the detected signal and
the error probability becomes equal to $\nu$. Another interesting
observation in Fig. \ref{fig:ooqpskriciannk5} is the existence of
error floors for sufficiently high values of SNR. This is due to the
fact that even if the additive noise vanishes, the performance is
limited by the multiplicative noise introduced by unknown fading.
Note that in the correct detection probability expressions
(\ref{eq:Peqn1}) and (\ref{eq:Peqn1T0}), the arguments of the
Q-function have the term $P$ in both the numerator and denominator.
Hence, letting $P \to \infty$ or $N_0 \to 0$ does not drive the
correct detection probabilities to 1.

\vspace{-.2cm}
\section{Conclusion} \label{sec:conclusion}

We have studied the error performance of peaky signaling over fading
channels. We have considered two modulation formats: OOBPSK and
OOQPSK. We have found the optimal MAP decision rules and obtained
analytical error probability expressions. Through numerical
examples, we have seen that error performance improves if the
peakedness of the signaling schemes are sufficiently increased. For
fixed error probabilities, substantial gains in terms of SNR per bit
are realized, making the peaky signaling schemes energy efficient
and hence desirable modulation formats for wireless sensor networks
where low-data rate transmissions with low energy consumption are
required \cite{Wang}.

%\vspace{-2cm}

\end{document}